\documentclass{PoS}

\title{Nuclear Force from Lattice QCD}

\ShortTitle{Nuclear Force from Lattice QCD}

\author{\speaker{Noriyoshi ISHII}\\
  Department of Physics, University of Tokyo, Tokyo 113--0033, JAPAN\\
  E-mail: \email{ishii@rarfaxp.riken.jp}
}
\author{Sinya AOKI\\
  Graduate School of Pure and Applied Science,
  University of Tsukuba, Tukuba 305--8571, Ibaraki, JAPAN\\
  E-mail: \email{saoki@het.ph.tsukuba.ac.jp}
}
\author{Tetsuo HATSUDA\\
  Department of Physics, University of Tokyo, Tokyo 113--0033, JAPAN\\
  E-mail: \email{hatsuda@phys.s.u-tokyo.ac.jp}
}

\newcommand{\Hs}{\hspace*{1em}}

\newcommand{\Eq}[1]{Eq.~(\ref{#1})}
\newcommand{\Fig}[1]{Fig.~(\ref{#1})}
\newcommand{\Ref}[1]{Ref.~\cite{#1}}
\newcommand{\alt}{\;\;\raisebox{0.5ex}{<}\hspace*{-0.7em}\raisebox{-0.5ex}{$\sim$}\;}
\newcommand{\agt}{\;\;\raisebox{0.5ex}{>}\hspace*{-0.7em}\raisebox{-0.5ex}{$\sim$}\;}

\abstract{  The  first lattice  QCD  result  on  the nuclear  force 
(the NN
potential) is  presented in the  quenched level.  The  standard Wilson
gauge action and the standard  Wilson quark action are employed on the
lattice of the size $16^3\times  24$ with the gauge coupling $\beta=5.7$
 and the hopping parameter $\kappa=0.1665$.
To  obtain the NN  potential, we  adopt a  method recently
proposed  by CP-PACS  collaboration to  study the  $\pi\pi$ scattering
phase shift.
It turns out that this  method provides the NN  potentials which
are faithful to those obtained in the analysis of NN scattering data.
By   identifying   the   equal-time Bethe-Salpeter   wave   function   with   the
Schr\"{o}dinger   wave  function   for the two nucleon system,
  the   NN  potential   is
reconstructed so that the  wave function satisfies the 
 time-independent Schr\"{o}dinger equation.
In this report, we restrict ourselves to  the $J^{P}=0^+$ and $I=1$ channel, which enables
us to pick up unambiguously the ``{\em central}'' NN potential $V_{\rm
central}(r)$.
The resulting  potential is seen to  posses a clear  repulsive core of
about 500 MeV at short distance ($r \alt 0.5$ fm).
Although  the  attraction  in the  intermediate  and  long
distance regions is still missing in the present lattice set-up, 
 our method is appeared to be quite
promising in reconstructing the NN potential with lattice QCD.  }

\FullConference{XXIVth International Symposium on Lattice Field Theory\\
                July 23-28, 2006\\
                Tucson, Arizona, USA}

\begin{document}
\section{Introduction}
The  nuclear  force (the  NN  potential) serves  as  one  of the  most
important building blocks in nuclear physics.
Its  attractive part  in  the intermediate  distance  region plays  an
essential role to bind nucleons  in nuclei.  On the other hand, its
strong repulsive core  at short distance ensures  the stability of
heavy  nuclei  by  preventing  them  from collapsing and  leads  to  the
celebrated saturation phenomena of nuclear matter.
Importance of the nuclear force goes even beyond the nuclear structure:
It can have a great influence  on the physics of compact stars and
the supernova explosion  through the equation of state at high baryon densities.

So far, enormous effort has been devoted to understand the nuclear force theoretically
\cite{oka}.  
The long  distance region  ($r \agt 2$ fm) 
of the  nuclear force is  dominated by the
exchange  of a  single pion,  i.e., the  lightest elementary
excitation in the QCD spectrum.
The intermediate  distance region ($1 \alt r \alt  2$ fm) receives  a significant contribution from multi-pion exchanges and from
 heavy  meson exchanges such as  $\rho$, $\omega$, and ``$\sigma$''.
The understanding of the   short
distance region ($r \alt 1$  fm)
is most retarded, but it is  believed to be 
 intimately related to the quark-gluon structure of the nucleons.

Hence, the first principle QCD calculation of the nuclear force
 (in particular the short distance region)
  has been desired for a long time.
Recently, a method attempting  to
obtain  the origin  of the repulsive  core on the lattice was proposed, where 
 the nucleon is assumed to be composed of  heavy-light-light quarks
  so that one can define the relative distance between the nucleons
  in terms of the distance between the heavy quarks
 \cite{rabbit}.  (The  same technique  was also used  in \Ref{mihaly}.)
In this report, we introduce a totally different  approach to 
 the NN  potential on the basis of a method recently
proposed  by   CP-PACS  collaboration  in  the   context  of  $\pi\pi$
scattering phase shift \cite{ishizuka}.
%
 For simplicity, we  confine ourselves to  the channel  $J^{P}=0^+$ and
$I=1$, which  makes us possible to obtain  unambiguously the so-called
{\it central} potential $V_{\rm central}(r)$.
First, we construct the equal-time Bethe-Salpeter (BS) wave function for the 
NN system. It is then
identified  with the  non-relativistic  Schr\"{o}dinger wave
function for two nucleons interacting at low energies.
  $V_{\rm central}(r)$ is reconstructed so that the wave function
 satisfies the non-relativistic Schr\"{o}dinger equation.
In this way, we do not have to
 introduce heavy quarks to define the relative distance 
 unlike the method proposed before.

The reconstructed  $V_{\rm central}(r)$ has a repulsive  core of about
0.5 GeV  in the short distance  region ($r \alt 0.5$  fm). On the other
 hand, the
attraction  in  the intermediate  and long  distance  regions is  still
missing, which  may be attributed  to combined artifacts of  the heavy
pion mass ($m_{\pi} \sim 0.5$  GeV), the small spatial volume ($L \sim
2.2$ fm) and statistics.

\section{The formalism}
We consider the equal-time Bethe-Salpeter (BS) wave function $\phi(\vec x; k)$ for
NN system in the $J^{P}=0^+$ and $I=1$ channel as
\begin{eqnarray}
  \phi(\vec x; k)
  &\equiv&
  \frac1{24}
  \sum_{R\in O}
  \frac1{L^3}
  \sum_{\vec X}
  (\sigma_2)_{\alpha\beta}
  \left\langle 0 \left|
  p_{\alpha}(R\cdot \vec x + \vec X)
  n_{\beta} (\vec X)
  \right|
  pn; k
  \right\rangle,
  \label{bs.wave.function}
  \\\nonumber
  p_{\alpha}(x)
  &\equiv&
  \epsilon_{abc}
  \left( u^T_a C\gamma_5 d_b \right)
  u_{c,\alpha},
  \\\nonumber
  n_{\beta}(x)
  &\equiv&
  \epsilon_{abc}
  \left( u^T_a C\gamma_5 d_b \right)
  d_{c,\beta},
\end{eqnarray}
where $a$, $b$ and $c$  denote the color indices, $\alpha$ and $\beta$
the   Dirac  indices,  and   $C\equiv  \gamma_4\gamma_2$   the  charge
conjugation   matrix.   $p_{\alpha}(x)$   and   $n_{\beta}(y)$  denote
composite  bispinor   fields  for   proton  and  neutron,   which  are
represented with Dirac's convention.
$\vec x$ plays  the role of the spatial separation  of the proton from
the neutron.
The summation over $R$ is performed for the cubic transformation group
$O$ to  pick up  the s-wave orbital  contribution of NN  system.  Note
that, due  to the parity  selection rule, only s-wave  contribution is
allowed for $J^{P}=0^+$.
The spinors of the proton and  the neutron are combined into scalar in
the  non-relativistic  manner  with $(\sigma_2)_{\alpha\beta}$.   
Since we are interested  in the significantly low energy region,
the lower  components of the  nucleon bispinors, i.e.,  $p_\alpha$ and
$n_\beta$, are negligible.
Since  this wave function  is symmetric in  orbit and
anti-symmetric  in  spin configuration,  it  has  to  be symmetric  in
iso-spin space due to the Pauli statistics, i.e., this wave function is
iso-vector.
The summation over $\vec X$  is performed to project the total spatial
momentum to zero, i.e., $\vec P = \vec 0$.
The L\"{u}scher's  $k$ is introduced as a  relative momentum outside
the range of the NN interaction, playing the role of ``{\it asymptotic
relative  momentum}''  between  the  proton  and the  neutron  in  the
scattering theory in the infinite volume limit.

The BS wave function \Eq{bs.wave.function} is obtained, in the lattice
QCD formulation, from the large $t$ behavior of the four point nucleon
correlator as
\begin{eqnarray}
  F_{NN}(\vec x,\vec y, t; t_0)
  &\equiv&
  \left\langle 0 \left|
  p_{\alpha}(\vec x,t)
  n_{\beta} (\vec y,t)
  \bar p_{\alpha'}(\vec 0,t_0)
  \bar n_{\beta'} (\vec 0,t_0)
  \right| 0 \right\rangle
  \\\nonumber
  &=&
  \sum_{n}
  \left\langle 0 \left|
  p_{\alpha}(\vec x)
  n_{\beta} (\vec y)
  \right| n \right\rangle\;
  A_n\;
  e^{-E_n(t - t_0)},
\end{eqnarray}
where $E_n$  denotes the  energy of the  state $|n\rangle$,  $t_0$ the
time-slice on  which the source is  located.  $A_n \equiv  \langle n |
\bar p_{\alpha}(0)  \bar n_{\beta}(0)|0\rangle$  plays the role  of an
``{\it overlap}''.
\Eq{bs.wave.function} satisfies the Bethe-Salpeter (BS) equation, since
it is obtained as  a solution to the BS equation for  the NN system in
the BS framework.
By  using the  procedure  of the  non-relativistic  reduction, the  BS
equation   reduces   to   the   effective   Schr\"{o}dinger   equation
\cite{luescher} as
\begin{equation}
  (\nabla^2 + k^2) \phi(\vec x; k)
  =
  m_N
  \int d^3 y\;
  V_k(\vec x - \vec y)
  \phi(\vec y; k),
  \label{effective.schroedinger.eq}
\end{equation}
where the  nucleon mass  $m_N$ is introduced  in the r.h.s.  for later
convenience.
The interaction kernel $V_k(\vec x  - \vec y)$, in general, depends on
$k$, which  is the  main reason why  \Eq{effective.schroedinger.eq} is
referred to as ``{\it effective}'' Schr\"{o}dinger equation.
At low  energies, $V_k$  can be  approximated as $V_k(\vec  x -  \vec y)
\simeq  V_{k\equiv  0}(\vec  x -  \vec  y)$.   We  are left  with  the
Schr\"{o}dinger equation for the NN system as
\begin{equation}
  -\frac1{2\mu} \nabla^2 \phi(\vec x; k)
  +
  \int d^3 y\;
  V_{NN}(\vec x - \vec y)
  \phi(\vec y; k)
  =
  E
  \phi(\vec x; k),
  \label{schrodinger.eq2}
\end{equation}
where $\mu\equiv m_N/2$  denotes the reduced mass of  the nucleon, and
$E\equiv k^2/(2\mu)$ denotes the non-relativistic ``{\it energy}''.

According to  the standard nuclear physics  \cite{ring-schuck}, the NN
interaction $V_{NN}$ is parameterized in  the low energy region as
\begin{eqnarray}
  V_{NN}
  &=&
  V_{\rm central}(r)
  + V_{\rm tensor}(r) \hat S_{12}
  + V_{\rm LS}(r) \hat{\vec L}\cdot \hat{\vec S}
  + O(p^2),
  \\\nonumber
  \hat S_{12}
  &\equiv&
  3
  { (\hat{\vec \sigma}_p \cdot \vec x)
    (\hat{\vec \sigma}_n \cdot \vec x)
    \over
    r^2
  }
  - (\hat{\vec \sigma}_p \cdot \hat{\vec \sigma}_n),
  \\\nonumber
  \hat{\vec L}
  &\equiv&
  -i \vec x \times \vec \nabla,
  \Hs\Hs
  \hat{\vec S}
  \equiv
  (\hat{\vec \sigma}_p + \hat{\vec \sigma}_n)/2,
\end{eqnarray}
where $r \equiv |\vec x|$  denotes the distance between the proton and
the  neutron.  $\hat{\vec\sigma}_p$  and $\hat{\vec  \sigma}_n$ denote
the   spin  Pauli   matrices   for  the   proton   and  the   neutron,
respectively. $\hat{\vec L}$  represents the relative angular momentum
operator  of the  proton from  the neutron.   $V_{\rm  central}(r)$ is
referred to as  the ``{\it central}'' NN force,  which only depends on
the distance  $r$.  It  is considered to  be the most  important local
interaction in the NN  interaction. (Interactions which do not involve
any  derivatives   are  referred  to  as   {\it  ``local''}.)  $V_{\rm
tensor}(r)$ is referred to as the ``{\it tensor}'' force, which is the
second important local interaction.  $V_{\rm LS}(r)$ is referred to as
the  ``{\it spin  orbit interaction}'',  which is  the  most important
non-local  term.   Finally,  $O(p^2)$ represents  remaining  non-local
contributions, which is often considered to be less important.

Restriction    to   the  $J^P=0^+$    channel    permits     a    further
simplification. Because  only the s-wave  orbital component is  allowed in
this   quantum  number  due   to  the   parity  selection   rule,  the
contributions  from the  second and  the third  terms  vanish.  Since,
then, $V_{NN}$  receives the contribution only from  the central term
as
\begin{equation}
  V_{NN}
  \simeq
  V_{\rm central}(r),
\end{equation}
\Eq{schrodinger.eq2} reduces to
\begin{equation}
  - \frac{\vec\nabla^2}{2\mu}
  \phi(\vec x; k)
  +
  V_{\rm central}(r)
  \phi(\vec x; k)
  =
  E
  \phi(\vec x; k).
\end{equation}
Since $V_{\rm  central}(r)$ is a simple multiplication
operator, we can rearrange it in the following way:
\begin{equation}
  V_{\rm central}(r)
  =
  E
  +
  \frac1{2\mu}
  {
    \vec\nabla^2 \phi(\vec x; k)
    \over
    \phi(\vec x; k)
  }.
  \label{CP-PACS-formula}
\end{equation}
This relation states that, by identifying the BS wave function for NN,
which   is   obtained   from   lattice  QCD   calculations,   as   the
Schr\"{o}dinger   wave   function,   $V_{\rm   central}(r)$   can   be
reconstructed so that the Schr\"{o}dinger equation is satisfied.
%
%
It should be emphasized that,  by construction, nothing is required on
the  quark mass  to  define  the fixed  separation between the two
nucleons. Since  the NN  potential is reconstructed  from the  NN wave
function, this method  is expected to provide such  NN potentials that
are faithful  to those obtained in  the analysis of  the NN scattering
data.

\section{The numerical calculation}
\begin{figure}
\begin{center}
\includegraphics[width=0.5\textwidth,angle=270]{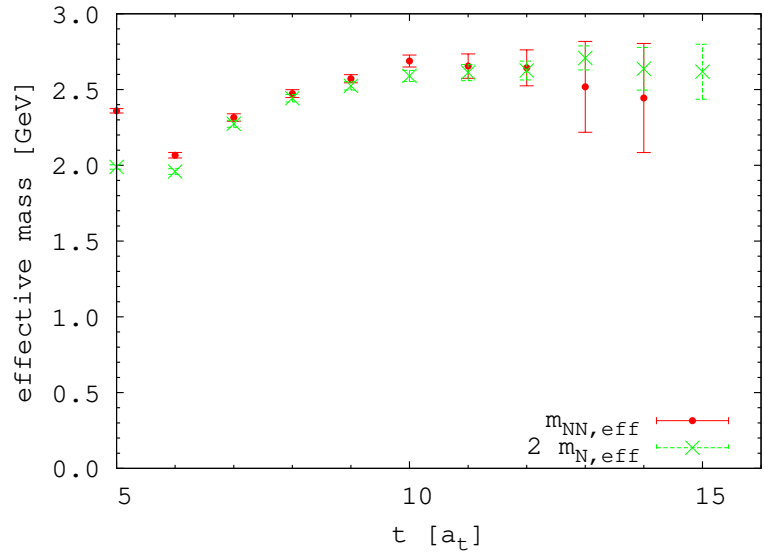}
\end{center}
\caption{The effective  mass plot. Dots  denote the effective  mass of
the  NN system ($J^P=0^+$,  $I=1$), whereas  crosses denote  twice the
effective mass of the nucleon.}
\label{effmass}
\end{figure}

We  employ the  standard Wilson  gauge  action at  the gauge  coupling
$\beta=5.7$ on the $16^3\times  24$ lattice together with the standard
Wilson quark  action with  the hopping parameter  $\kappa=0.1665$. The
lattice spacing  is determined  from $\rho$ meson  mass in  the chiral
limit  as $a^{-1}=1.44(2)$ GeV  ($a \simeq  0.137$ fm).   The physical
volume      of     our     lattice      is     $(2.2\mbox{fm})^3\times
(3.3\mbox{fm})$. These parameters reproduces $m_{\pi}\simeq 0.53$ GeV,
$m_{\rho}\simeq  0.88$   GeV,  and  $m_{N}  \simeq   1.3$  GeV.   (See
Ref.~\cite{fukugita}  for detail.) The  global heat-bath  algorithm is
used to generate the gauge configurations.  After skipping 5000 sweeps
for thermalization,  the gauge configurations are picked  up every 100
sweeps.   Totally,   160  gauge   configurations  are  used   for  the
measurement.
Dirichlet boundary condition is imposed  on the quark fields along the
temporal  direction on the  time-slice $t=0$.   To enhance  the ground
state  contribution of  the NN  wave  function, we  adopt a  spatially
extended  source  with the  gaussian  smearing  method  with the  size
$\rho\simeq 0.4$ fm on the time-slice $t=5$.  The BS wave function for
NN is  measured on the time-slice  $t=10$, which is  determined from a
plateau   appearing  in   the  NN   effective  mass   plot   shown  in
\Fig{effmass}.  We  keep in mind  that the ground state  saturation of
the BS wave function is quite important.

\begin{figure}
\begin{center}
\includegraphics[width=0.5\textwidth,angle=270]{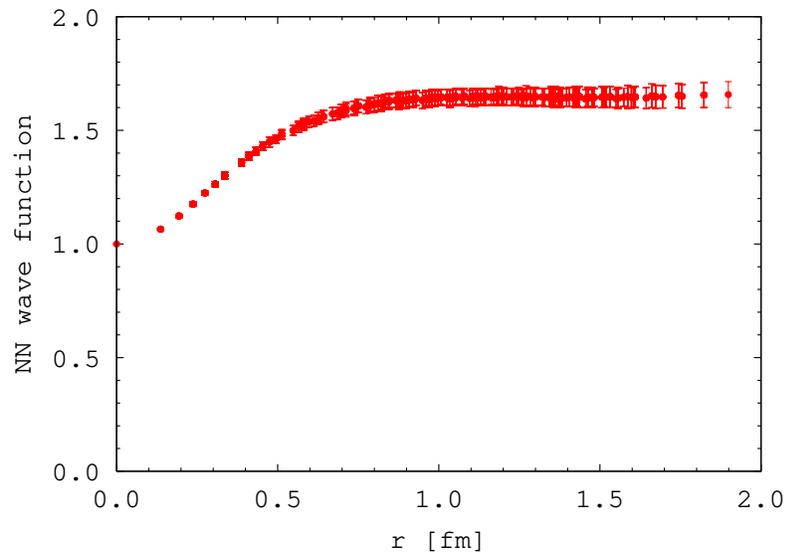}
\end{center}
\caption{The lattice  QCD result of  the NN wave  function ($J^P=0^+$,
$I=1$).}
\label{wave-function}
\end{figure}
\Fig{wave-function}  shows  the lattice  QCD  result  of  the BS  wave
function for NN. It is normalized with its value at the origin $\vec r
= \vec 0$. Not only the on-axis  data , but also the off-axis data are
plotted all together, due to  which the horizontal axis extends beyond
$L/2 \sim  1.1$ fm to $\sqrt{3}L/2  \sim 1.9$ fm ($L\simeq  2.2$ fm is
the spatial lattice extension).
We observe that  the wave function shrinks near  the origin suggesting
the existence of repulsion at short distance.

\begin{figure}
\begin{center}
\includegraphics[width=0.5\textwidth,angle=270]{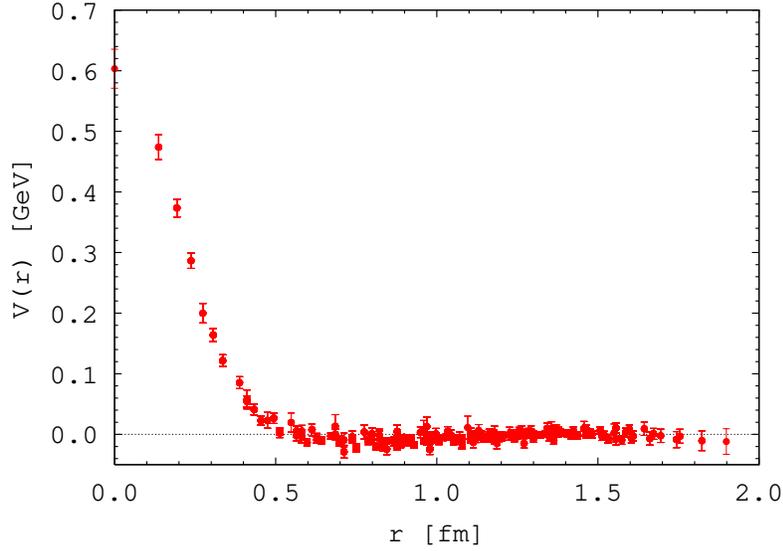}
\end{center}
\caption{The  lattice   QCD  result   of  the  NN   potential  $V_{\rm
central}(r)$.}
\label{NN-potential}
\end{figure}
\Fig{NN-potential} shows  the lattice QCD  result of the  NN potential
$V_{\rm central}(r)$ for $J^P=0^+$, $I=1$ channel.
The    simplest    nearest    neighbor   numerical    Laplacian    for
\Eq{CP-PACS-formula}    is   employed   in    reconstructing   $V_{\rm
center}(r)$.
The   zero   adjustment    due   to   $E\equiv   \frac1{2\mu}k^2$   in
\Eq{CP-PACS-formula} has  not yet been performed because  of the large
noise.   In principle,  L\"{u}scher's  $k$ can  be  obtained from  the
difference of the effective masses, i.e., $m_{NN,\rm eff} - 2 m_{N,\rm
eff}$ (cf.  \Fig{effmass}),  or the NN wave function  in the large $r$
region \cite{ishizuka}.
So far, the data is too noisy to obtain it.
We observe  that, corresponding to the  shrink in \Fig{wave-function},
$V_{\rm central}(r)$  has a clear repulsive  core of about  500 MeV in
the short  distance region $r \alt 0.5$  fm. On the other  hand, we do
not find a significant attraction in the intermediate distance region,
which  may be attributed  to (i)  the still  large noise,  (ii) finite
volume effect  ($L\sim 2.2$ fm is  too small for NN),  (iii) the large
pion mass ($m_{\pi} \sim 0.53$ GeV).
In order  to gain  physics insights into  the origin of  the repulsive
core at  short distance,  we need more  information such as  the quark
mass dependence, the flavor dependence, etc.

\section{Summary and discussion}
 To study the nuclear force, we  have applied  a new  method 
recently proposed by CP-PACS collaboration in the
context of the $\pi\pi$ scattering phase shift.
In this method, constraint on the quark mass to define
the distance between the two nucleons is not necessary.
Because the NN  potential is reconstructed from the  NN wave function,
it is expected  to provide NN potentials, which  are faithful to those
obtained from the analysis of the NN scattering data.
By restricting ourselves to the quantum number $J^P=0^+$ and $I=1$, we
have  reconstructed  the  central  part of  the  NN-potential  $V_{\rm
central}(r)$  from the  Bethe-Salpeter wave  function for  NN obtained
by using lattice QCD.
We have seen  that $V_{\rm central}(r)$ has a  clear repulsive core of
about $500$ MeV in the short distance region $r \alt 0.5$ fm. However,
we  have  not  found  a  significant attraction  in  the  intermediate
distance region, which  may be attributed to the  poor statistics, the
small  lattice  volume ($L  \sim  2.2$ fm)  and  the  large pion  mass
$m_{\pi} \sim  0.53$ GeV.  We  are currently performing a  lattice QCD
calculation for  $V_{\rm central}(r)$ using  a better statistics  on a
larger  lattice  volume  with  a  lighter pion  mass,  which  will  be
presented elsewhere.
Our method have  appeared to be quite promising  in reproducing the NN
potential with lattice QCD.
It is interesting  to apply our method to  other channels to construct
$V_{\rm LS}(r)$ and $V_{\rm tensor}(r)$.  Also, it is important to apply our
method  to hyperon-nucleon and hyperon-hyperon
 forces, where only  a limited
number of experimental data are available so far.

\begin{center} Acknowledgments \end{center}
This research  was partially supported  by the Ministry  of Education,
Science, Sports  and Culture, Grant-in-Aid for  Scientific Research on
Priority Areas No.~15540251  and No.~17070002, Scientific Research (C)
No.~13135204 and No.~18540253.
Lattice QCD Monte Carlo calculations have been performed with IBM Blue
Gene/L at KEK under a support of its Large Scale Simulation Program.

\end{document}